\documentclass[preprint,12pt]{elsarticle}

\usepackage{amssymb}
\usepackage{amsmath}
\usepackage{tikz}
\usepackage{amsmath}  % improve math presentation
\usepackage{graphicx}
\usepackage{nomencl}
\usepackage{subfigure}
\usepackage{xcolor}
\usepackage{color, soul}
\usepackage{multirow}
\usepackage{tabularx}

\makenomenclature

\begin{document}

\begin{frontmatter}

\title{Numerical Simulation of Turbulent Concentric Annular Pipe Flow using One-Dimensional Turbulence (ODT): Part 1: Momentum Transfer}

%% use optional labels to link authors explicitly to addresses:
\author[add1,add2]{Pei-Yun Tsai}
\affiliation[add1]{
            % organization={Chair of numerical },
            addressline={Chair of Numerical Fluid and Gas Dynamics, Brandenburg University of Technology Cottbus-Senftenberg, Siemens-Halske-Ring 15A, 03046 Cottbus, Germany}
            }

\affiliation[add2]{
            organization={Scientific Computing Lab (SCL), Energy Innovation Center (EIZ), Brandenburg University of Technology Cottbus-Senftenberg},
            addressline={03046 Cottbus, Germany}
            }

\author[add1,add2]{Marten Klein}

\author[add1,add2]{Heiko Schmidt}

%%%%%%%%%%%%%%%%%%%%%%%%%%%%%%%%%%%%%%%%%%%%%%%%%%%%%%%%%%%%%%%%%%%%%%%%%%%%%%%
\begin{abstract}
Turbulent concentric coaxial (annular) pipe flow is numerically investigated using a stochastic one-dimensional turbulence (ODT) model as a stand-alone tool. The dimensionally reduced ODT domain enables fully resolved numerical simulations of the flow across the radial gap between the cylindrical inner wall and the cylindrical outer wall. The model is calibrated with available reference data at low bulk Reynolds number $Re_{D_h} = 8900$ for a wide (radius ratio $\eta = 0.1$) and a moderate ($\eta = 0.5$) gap. Making use of the model's predictive capabilities, radius ratio and Reynolds number effects are investigated, reaching bulk Reynolds numbers as large as $Re_{D_h}=10^6$. Despite the large $Re_{D_h}$ values reached, spanwise wall-curvature effects remain sensible in the momentum boundary layer. The effects are more pronounced for larger wall curvature and to leading orders restricted to the convex cylindrical inner wall. Wall-curvature corrections to the law of the wall are obtained for both the viscous and Reynolds-stress dominated regions by fitting analytically derived expressions for the flow profile to the stochastic simulation data, demonstrating physical compatibility with Reynolds-averaged Navier--Stokes flow. Second-order and detailed fluctuation statistics demonstrate the permeating and nonlocal influence of spanwise wall curvature on the turbulent boundary layer. Surrogate model output in terms of conditional eddy event statistics reveals that the disparity between the near-inner and near-outer wall turbulence increases with Reynolds number for small radius ratios, suggesting that annular pipe flows require wall-curvature-aware wall models even at very large Reynolds numbers.

\end{abstract}

%%Graphical abstract
\begin{graphicalabstract}
\includegraphics[width=1.0\textwidth]{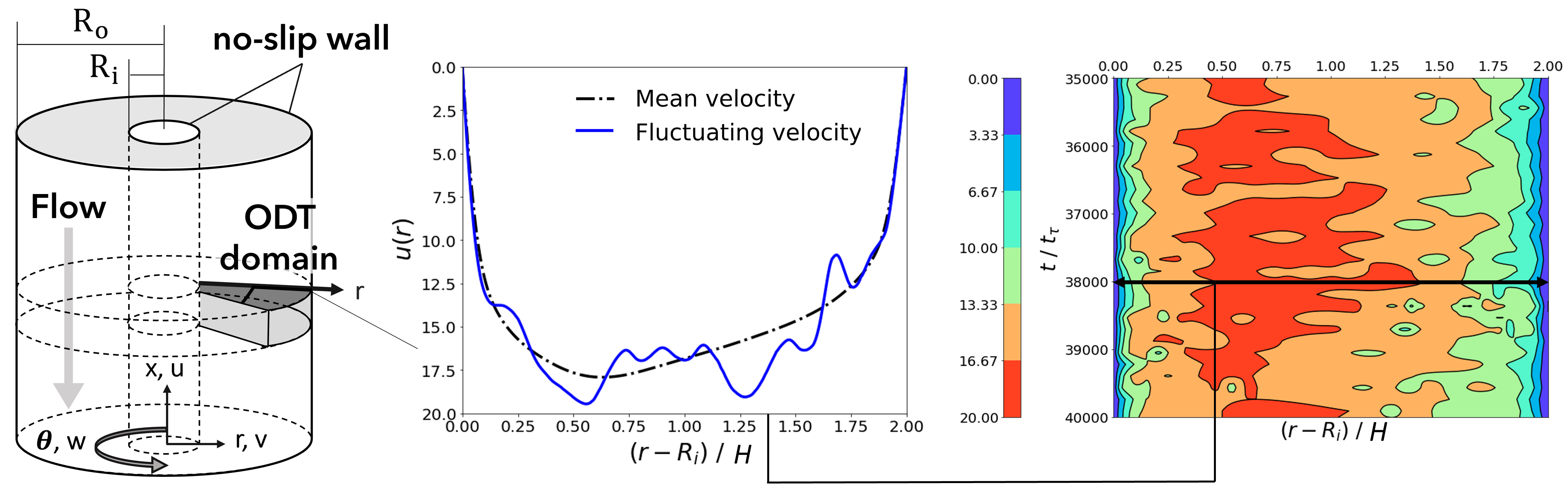}
\end{graphicalabstract}

%%Research highlights
\begin{highlights}
\item A stochastic one-dimensional turbulence model is applied to annular pipe flow.
\item Spanwise wall-curvature effects are investigated.
\item Radius ratio and Reynolds number are varied in an extended range of values.
\item An analytical expression of the boundary layer profile, including curvature effects, is given.
\item Wall-curvature effects on low-order and detailed turbulence statistics are reported and analyzed. 
\end{highlights}

%% Keywords

\end{frontmatter}

%% main text
%%
%%%%%%%%%%%%%%%%%%%%%%%%%%%%%%%%%%%%%%%%%%%%%%%%%%%%%%%%%%%%%%%%%%%%%%%%%%%
\section{Introduction}\label{sec:introduction}
%%%%% opening %%%%%
Turbulent annular pipe flow is of interest not only due to its direct engineering applications — such as coaxial geothermal heat exchangers~\cite{raymond2015,renaud2021,chen-tomac2023}, photochemical reactors~\cite{balestrin2021}, and gas-cleaning devices~\cite{shapiro1989,mendez2022} — but also because it offers valuable physical insights into the general problem of fully developed turbulent shear flow. Different from the plane-channel flow and circular pipe flow, the momentum transport in the concentric annular pipe flow is significantly influenced by the transverse curvatures from the inner and outer cylindrical walls. The radius ratio $\eta=R_\text{i}/R_\text{o}$ in the annular pipe, where $R_\text{i}$ and $R_\text{o}$ are the radii of the inner and outer pipes, respectively, introduces asymmetry in the mean velocity properties and turbulence structures near the two curved walls. While the curvature effect becomes nearly negligible at very high Reynolds number flow, it remains apparent in both first-order and second-order flow statistics under finite-Reynolds-number conditions, making the analysis of this annular pipe flow more challenging. In fact, plane channel flow and circular pipe flow can be considered as limiting cases of the annular pipe flow, corresponding to radius ratio $\eta\rightarrow1$ and $\eta\rightarrow0$, respectively.  

%%%%% literature reviews %%%%% 
In recent decades, numerous numerical and experimental studies have been carried out to better understand the characteristics of pressure-driven (Poiseuille) turbulent flow in concentric annular pipes. In the last century, a major debate centered on whether the location of maximum radial velocity coincides with the point of zero shear stress. Several experimental studies were conducted on this topic~\cite{brighton1964, quarmby1967experimental, rehme1974, hernandez2019piv}. Brighton and Jones (1964)~\cite{brighton1964} and Quarmby (1967)~\cite{quarmby1967experimental} found that in turbulent flow within concentric annuli, the radial position of maximum axial velocity coincides with the location of zero shear stress. The latter also found that this position is closer to the inner pipe wall in turbulent flow compared to laminar flow, with the difference increasing as the radius ratio decreases. On the other hand, Rehme (1974)~\cite{rehme1974} experimentally observed that the location of zero shear stress does not coincide with the point of maximum axial velocity. Specifically, he found that the zero shear stress position is closer to the inner pipe wall compared to the location of maximum axial velocity. Nouri \textit{et al.} (1993)~\cite{nouri1993} performed a laser-Doppler velocimetry (LDV) experiment in concentric annular pipes for both Newtonian and non-Newtonian fluids at radius ratio $\eta=0.5$. Chung \textit{et al.} (2002)~\cite{chung2002} conducted a comprehensive direct numerical simulation (DNS) study of turbulent flow in concentric annular pipes with radius ratios $\eta=0.1$ and $0.5$, maintaining a bulk Reynolds number $Re_{D_h}=8900$, consistent with the experimental conditions of Nouri \textit{et al.} (1993)~\cite{nouri1993}. Here, the bulk Reynolds number is defined as $Re_{D_h}=\overline{u}_\text{b}\,D_{h}\,/\,\nu$, where $\overline{u}_\text{b}$ denotes the bulk mean velocity, $D_h=2(R_\text{o}-R_\text{i})$ represents the hydraulic diameter, which is twice the gap width between outer $R_\text{o}$ and inner $R_\text{i}$ pipe radius, and $\nu$ is the kinematic viscosity of the fluid. Boersma and Breugem (2011)~\cite{boersma2011} performed DNS studies on turbulent flow in annular pipes with small radius ratios ($\eta=0.02, 0.04$ and $0.1$) across a moderate range of Reynolds numbers. Ishida \textit{et al.} (2016, 2017)~\cite{ishida2016friction, ishida2017turbulent} carried out a large-scale DNS of pressure-driven annular pipe flows, investigating radius ratio ranging from $\eta=0.1$ to $\eta=0.8$ with the subcritical transitional regime at friction Reynolds number $Re_\tau$ between $48$ and $150$. The results indicated that the shape of transitional structures varies with the radius ratio $\eta$, with helical turbulence structures becoming dominant for $\eta\geq 0.5$. Hern{\'a}ndez-Cely \textit{et al.} (2019)~\cite{hernandez2019piv} conducted experimental measurement through particle image velocimetry (PIV) to investigate the influence of radius ratio and Reynolds number on the radial location of the maximum mean axial velocity. From their observation, they found that it is mainly affected by the radius ratio while almost not influenced by the Reynolds number. More recently, Bagheri and Wang (2020)~\cite{bagheri2020} investigated the influence of radius ratio on the turbulent statistics and structures in both physical and spectral spaces using DNS for multiple radius ratio cases ($\eta=0.1,0.3,0.5$ and $0.7$) with bulk Reynolds number $Re_{D_h}=8900$. As an extension, turbulent heat transfer in the annular pipe flow, coupled with a passive scalar, has also been discussed in previous studies~\cite{chung2003,bagheri2021,wang2025}. 

The above literature review indicates that in annular pipe flows, the curvature difference between the inner and outer cylindrical walls significantly influences the distribution of velocity, shear stress, and turbulence characteristics. However, numerical studies on annular pipe flows at high Reynolds numbers~\cite{tsai2025,tsai2023,klein2022stab} remain limited due to the high-resolution demands of high-fidelity simulations~\cite{boersma2011}. Additionally, a theoretical explanation for empirical pressure loss properties remains challenging, as direct measurement of the boundary layer near the curved inner wall is difficult. Therefore, this study aims to contribute to the development of improved wall functions for spanwise curved walls, enabling their application in classical large-eddy and Reynolds-averaged simulations. To achieve this, we adopt an alternative approach utilizing Kerstein's one-dimensional turbulence (ODT) model~\cite{kerstein1999,kerstein2001}, which captures the evolution of instantaneous boundary layer profiles. 

%%%%% paper structure %%%%%
The rest of this paper is organized as follows. Section~\ref{sec:formulation} gives the details of the annular pipe flow setup investigated and provides an overview of the ODT model formulation and the governing equations. Section~\ref{sec:results} discusses the main results in terms of velocity statistics, focusing on examining the impact of spanwise curvature and Reynolds number on some turbulence properties. Last, Section~\ref{sec:conclusions} summarizes the key findings of this work.

%%%%%%%%%%%%%%%%%%%%%%%%%%%%%%%%%%%%%%%%%%%%%%%%%%%%%%%%%%%%%%%%%%%%%%%%%%%%%
\section{Model formulation}\label{sec:formulation}
%%%%% ODT background and case study %%%%%
The one-dimensional turbulence (ODT) model, introduced by Kerstein (1999)~\cite{kerstein1999}, offers a computationally efficient approach to simulating turbulent flows with full-scale resolution up to very high Reynolds numbers along a single spatial coordinate. Unlike conventional turbulence models, ODT represents turbulent advection through a stochastic point process, implemented via spatial triplet maps acting on the one-dimensional ODT domain~\cite{kerstein1999,kerstein2001,lignell2013}. Each stochastic event, termed an \textit{eddy event}, is characterized by three randomly sampled numbers: eddy size, eddy location, and time of occurrence, each of them from an unknown distribution. As an alternative to constructing a distribution function, rejection sampling is utilized, formulated with an instantaneous rate that depends on the current flow state \cite{kerstein1999,kerstein2001}.

Lignell \textit{et al.} (2018)~\cite{lignell2018} extended the original planar ODT formulation to cylindrical and spherical geometries, thereby enabling simulations of radial ODT domains, which is a crucial capability for annular pipes in the focus of this study. While the one-dimensional nature of ODT restricts its standalone application to simple geometries, it is well-suited for a number of relevant canonical turbulent shear flows, including boundary layers, channel flows~\cite{marten2022}, pipe flows~\cite{mendez2022}, and jet flows~\cite{marten2023jet} that are fundamental to turbulence research.

\begin{figure}[tbp]
    \centering
    \includegraphics[width=1.0\textwidth]{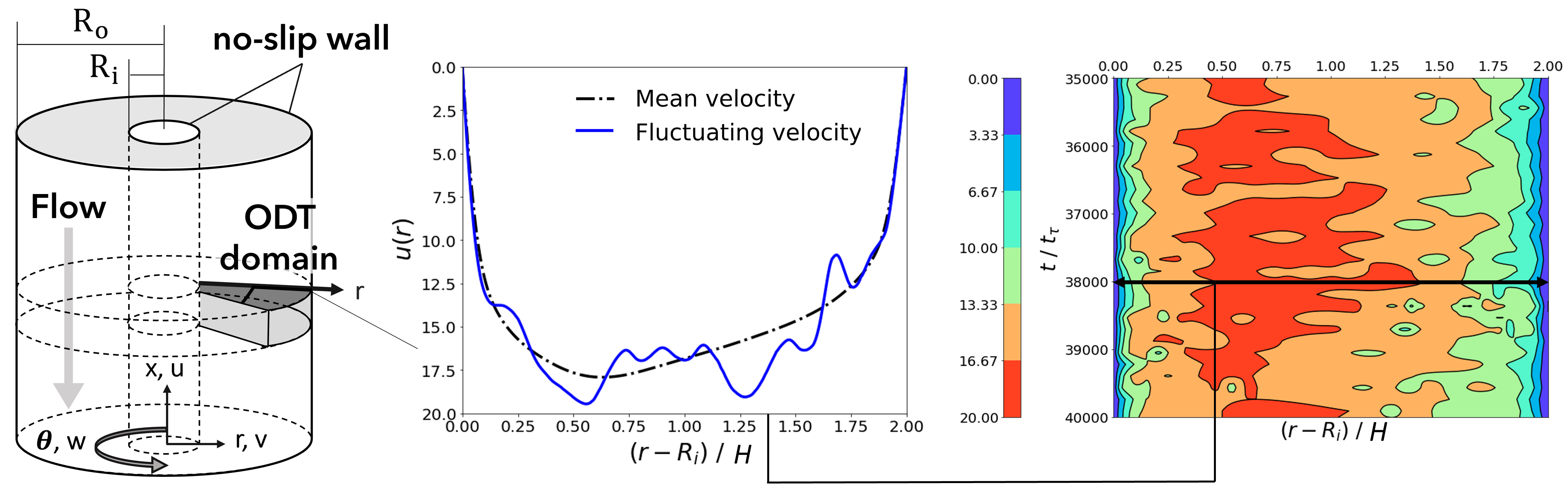}
    \caption{(\textit{left}) Schematic of the investigated concentric annular pipe flow configuration. No-slip boundary conditions are prescribed at the pipe walls, and an axial mean pressure gradient force is imposed to drive the flow. The numerical domain of the stochastic ODT simulations (ODT domain) represents a single radial coordinate spanning the entire gap between the cylinders. (\textit{middle}) The instantaneous and mean radial profile of the axial velocity component. (\textit{right}) Space-time diagram of a turbulent ODT solution for an annular pipe flow. $t$ and $t_\tau$ denote the simulation time and its time scale that is used for normalization. $H$ is half of the width between inner and outer cylinder walls, $H=(R_\text{o}-R_\text{i})/2$.
    }
    \label{fig:geometry}
\end{figure}

Here, we consider an incompressible, constant-property Poiseuille flow confined between two concentric, coaxial cylinders as sketched in Figure~\ref{fig:geometry}(\textit{left}). No-slip boundary conditions are applied at both the inner and outer cylinder walls, and an axial mean pressure gradient drives the flow. The ODT domain is oriented in the radial direction, spanning the gap between the cylinders. Figure~\ref{fig:geometry}(\textit{middle}, \textit{right}) shows a representative instantaneous axial velocity profile generated by the ODT simulation using the scale-locality preserving radial triplet map TMB~\cite{lignell2018}, alongside the mean velocity profile and a space-time diagram that captures the evolution of the velocity field.

%%%%% ODT model governing equation %%%%%
Following the notation introduced by Klein \textit{et al.}~\cite{marten2022}, the dimensionally reduced ODT governing equation for momentum in turbulent annular pipe flow can be written as 
\begin{equation}
   \frac{\partial {u}_i}{\partial t} 
   +\sum_{t_\text{e}}\mathcal{E}_i\,\delta\left(t-t_\text{e}\right)
   = \frac{1}{r}\frac{\partial}{\partial r}\bigg(r\nu\frac{\partial {u}_i}{\partial r}\bigg)
    -\frac{1}{\rho}\frac{{\rm d}P}{{\rm d}x}\,\delta_{ix}  \;\;.
    \label{eq:gov_vel}
\end{equation}
Here, ${u}_i$, $i=x,r,\theta$, represents the model-resolved instantaneous velocity vector in cylindrical coordinates, where $u_x=u$, $u_r=v$, and $u_\theta=w$ correspond to the axial, radial, and azimuthal directions, respectively. The variable $t$ denotes the time, while $t_\text{e}$ represents the stochastically sampled times of eddy event, $\mathcal{E}_i$ ($i=x,r,\theta$), occurrences. The function $\delta(t)$ is the Dirac distribution function. The term ${\text{d}P}/{\text{d}x}$ represents the prescribed mean pressure-gradient force in the axial direction, $\delta_{ij}$ is the Kronecker delta, $\nu$ denotes the kinematic viscosity, and $\rho$ is the mass density of the fluid. 

Turbulent advection in ODT is implemented through spatial mapping using the triplet map, where we adopted TMB~\cite{lignell2018} in this case. During each eddy event, the velocity components are changed according to the following symbolic form, as described in~\cite{lignell2018} that 
\begin{equation}
    u_i(f(r)) \rightarrow u_i^\prime(f(r)) + c_i(\alpha) K(f(r)) + b_i(\alpha) J(f(r))
    \; \; ,
    \label{eq:eddy_event}
\end{equation}
where $u_i(f(r))$ and $u_i^\prime(f(r))$ represent the pre-triplet map and post-triplet map velocity components, respectively. $K(r)$ and $J(r)$ are kernel functions, $K(r)$ is defined as the displacement of a point from its initial position by a triplet map and $J(r)=|K(r)|$. The coefficients $c_i(\alpha)$ and $b_i(\alpha)$ ensure conservation of momentum and energy in the model extension to cylindrical coordinates, and depend on the energy redistribution parameter $\alpha$, which governs the transfer of energy between velocity components in response to pressure fluctuations~\cite{kerstein2001}. For heated planar Poiseuille flow, values in the range $0 < \alpha < 2/3$ are typically recommended~\cite{marten2022}. In the present study, we use $\alpha = 1/6$ consistent with the model calibration for heated channels \cite{marten2022}, but the selected value is of minor influence compared to the primary model parameters $C$ and $Z$ that are discussed below.

Each eddy event is defined by its size, location, and occurrence time. In the case of annular pipe flow, the maximum eddy size is specified to reflect the transition between channel and pipe flow geometries. It is given as 
% \begin{equation}
%     L_{\text{max}}=\frac{1}{2}\eta+2\times\frac{1}{3}\left( 1-\eta \right)
%     \; \; ,
%     \label{eq:Lmax}
% \end{equation}
\begin{equation}
    L_{\text{max}}=\;\left( R_o-R_i \right) \left(\frac{1}{4}\eta+\times\frac{1}{3}\left( 1-\eta \right)\right)
    \; \; ,
    \label{eq:Lmax}
\end{equation}
where $L_{\text{max}}=L_{\text{max}}(\eta)$ for annular pipe flows results from a linear interpolation between the previously calibrated values of the maximum permissible eddy size in channel~\cite{marten2022} and pipe~\cite{lignell2018} flow, respectively. 

Eddy events are sampled based on an effective rate that reflects the instantaneous local flow condition. Building on a local interpretation of Prandtl’s mixing-length hypothesis, the original formulation by Kerstein~\cite{kerstein1999} was generalized in~\cite{kerstein2001} to an integral form that incorporates the eddy-integrated available energy. For turbulent shear flows in cylindrical coordinates, the eddy sampling frequency is generalized to~\cite{lignell2018} 
\begin{equation}
    \frac{1}{\tau_\text{e}} = C \sqrt{ \frac{2}{\rho_{KK} \, V_\text{e} \, l^2} \left( \frac{K_K}{V_\text{e} \, l^2} E_\text{kin} - Z\,E_\text{vp} \right) }
    \; \; ,
    \label{eq:para_odt_tau_e}
\end{equation}
where $\tau_\text{e}$ is the eddy time scale, $C$ and $Z$ are the adjustable ODT model parameters controlling the eddy rate and viscous penalty, respectively. In this case, we use $C=5$, and $Z=400$. $E_\text{kin}$ represents the kinetic energy, and $ZE_\text{vp}$ denotes the viscous penalty energy that is subtracted from $E_\text{kin}$. The eddy size and volume are given by $l$ and $V_\text{e}$, respectively. $K(r)$ is a kernel function, $K_K=\int_V K^2 dV$ is the map-induced fluid displacement, and $\rho_{KK} = \rho \, K_K$ is the kernel-weighted density, which is here simplified by limiting the attention to constant-property flow, as described in~\cite{lignell2018,marten2022}.

A detailed calibration of the model parameters $\alpha$, $C$, and $Z$ for heated annular pipe flow is available in~\cite{tsai2022}, including a demonstration of the predictive capabilities and a discussion of the model limitations. The ODT results are validated against DNS data from~\cite{chung2002,bagheri2020}, as discussed in Section~\ref{sec:results}. Simulation parameters and bulk flow quantities for the cases studied are summarized in Table~\ref{table:details}.

Further details of the ODT model formulation and its implementation are not repeated here. Instead, the reader is referred to the original conceptual development in~\cite{kerstein1999,kerstein2001} and to~\cite{lignell2018} for specifics on the fully adaptive cylindrical coordinate solver. The current work builds upon the open-source, object-oriented C++ ODT framework developed by Stephens and Lignell~\cite{stephens2021}.

%%%%% definition for friction quantities %%%%%
For the normalization purpose, the local friction velocities at the inner and outer walls are defined as 
\begin{equation}
    u_{\tau,\,\text{i}} = \sqrt{ \nu\left|\frac{\mathrm{d}u}{\mathrm{d}r}\right|_\text{wall,\,i} }
    \; \quad\text{and}\quad
    u_{\tau,\, \text{o}} = \sqrt{ \nu\left|\frac{\mathrm{d}u}{\mathrm{d}r}\right|_\text{wall,\,o} } 
     \; \; ,
    \label{eq:u_tau}
\end{equation}
where $u_{\tau,\,\text{i}}$ and $u_{\tau,\,\text{o}}$ denote the friction velocity on the inner and outer pipe walls, respectively. The corresponding friction Reynolds numbers are defined as $Re_{\tau,\,\text{i}}=u_{\tau,\,\text{i}}\,\delta_{t,\,\text{i}}\,/\,\nu$ and $Re_{\tau,\,\text{o}}=u_{\tau,\,\text{o}}\,\delta_{t,\,\text{o}}\,/\,\nu$, where $\delta_{t,\,\text{i}}$ and $\delta_{t,\,\text{o}}$ represent the boundary layer thicknesses at the inner and outer walls. The normalized mean velocity is given by $\overline{u}^+=u/u_{\tau,\,\text{o/i}}$ and the corresponding normalized radial coordinate is defined as $r^+=|r-R_\text{o/i}|\,u_{\tau,\,\text{o/i}}\,/\,\nu$. 

%%%%% detailed information for ODT simulation %%%%%
\begin{table}[tbp]
% \vspace{-2cm}
\centering
\caption{Details of ODT bulk quantities and simulation parameters for test cases with different radius ratio $\eta$ and bulk Reynolds number $Re_{D_h}$. $\overline{N_r}$ is the average number of cells in the adaptive grid, and $\Delta {r_\text{min}}^+$ is the normalized minimum allowed cell size.}
\label{table:details}
\begin{tabularx}{1\textwidth}{cccccccc}
\hline 
$Re_{D_h}$ & $\eta$ & $\delta_{t,\,\text{i}}\,/\,H$ & $\delta_{t,\,\text{o}}\,/\,H$ & $Re_{\tau,\,\text{i}}$ & $Re_{\tau,\,\text{o}}$ & $\overline{N_r}$ & $\Delta {r_\text{min}}^+$ \\ 
\hline
$8900$  &  $0.1$   & $0.615$ & $1.385$ & $110.0$  & $199.9$   & 319  & 0.165   \\
$8900$  &  $0.3$   & $0.803$ & $1.197$ & $130.8$  & $174.9$   & 326  & 0.168   \\
$8900$  &  $0.5$   & $0.895$ & $1.105$ & $140.8$  & $162.9$   & 329  & 0.165   \\ 
$8900$  &  $0.7$   & $0.947$ & $1.053$ & $145.8$  & $156.6$   & 327  & 0.165   \\ 
\hline
$10^5$  &  $0.02$  & $0.369$ & $1.631$ & $587.0$  & $1865.0$  & 891  & 0.153   \\
$10^5$  &  $0.04$  & $0.463$ & $1.537$ & $687.5$  & $1758.1$  & 925  & 0.154   \\
$10^5$  &  $0.06$  & $0.520$ & $1.480$ & $744.8$  & $1697.5$  & 916  & 0.155   \\
$10^5$  &  $0.08$  & $0.567$ & $1.433$ & $792.3$  & $1643.6$  & 983  & 0.155   \\
$10^5$  &  $0.1$   & $0.600$ & $1.400$ & $822.8$  & $1607.0$  & 942  & 0.155   \\
$10^5$  &  $0.3$   & $0.790$ & $1.210$ & $1001.5$ & $1401.8$  & 802  & 0.157   \\
$10^5$  &  $0.5$   & $0.882$ & $1.118$ & $1086.6$ & $1305.4$  & 940  & 0.158   \\ 
$10^5$  &  $0.7$   & $0.941$ & $1.059$ & $1139.8$ & $1245.2$  & 950  & 0.158   \\ 
$10^5$  &  $0.9$   & $0.988$ & $1.012$ & $1184.1$ & $1197.3$  & 855  & 0.158   \\
\hline
$10^6$  &  $0.02$  & $0.359$ & $1.641$ & $4229.5$ & $14727.2$ & 4768 & 0.158   \\
$10^6$  &  $0.04$  & $0.435$ & $1.565$ & $4847.6$ & $13981.2$ & 4787 & 0.158   \\
$10^6$  &  $0.1$   & $0.577$ & $1.423$ & $5987.4$ & $12773.4$ & 4843 & 0.159   \\
$10^6$  &  $0.5$   & $0.844$ & $1.156$ & $8044.8$ & $10492.7$ & 4896 & 0.161   \\ 
$10^6$  &  $0.7$   & $0.951$ & $1.049$ & $8921.3$ & $9560.2$  & 4871 & 0.161   \\ 
\hline   
\end{tabularx}
\end{table}

%%%%%%%%%%%%%%%%%%%%%%%%%%%%%%%%%%%%%%%%%%%%%%%%%%%%%%%%%%%%%%%%%%%%%%%%%%%%
\section{Results and discussion}\label{sec:results}
%%%%%%%%%%%%%%%%%%%%%%%%%%%%%%%%%%%%%%%%%%%%%%%%%%%%%%%%%%%%%%%%%%%%%%%%%%%%
\subsection{Mean velocity profiles}
\begin{figure}[tbp]
    \centering 
    \includegraphics[width=1.0\textwidth]{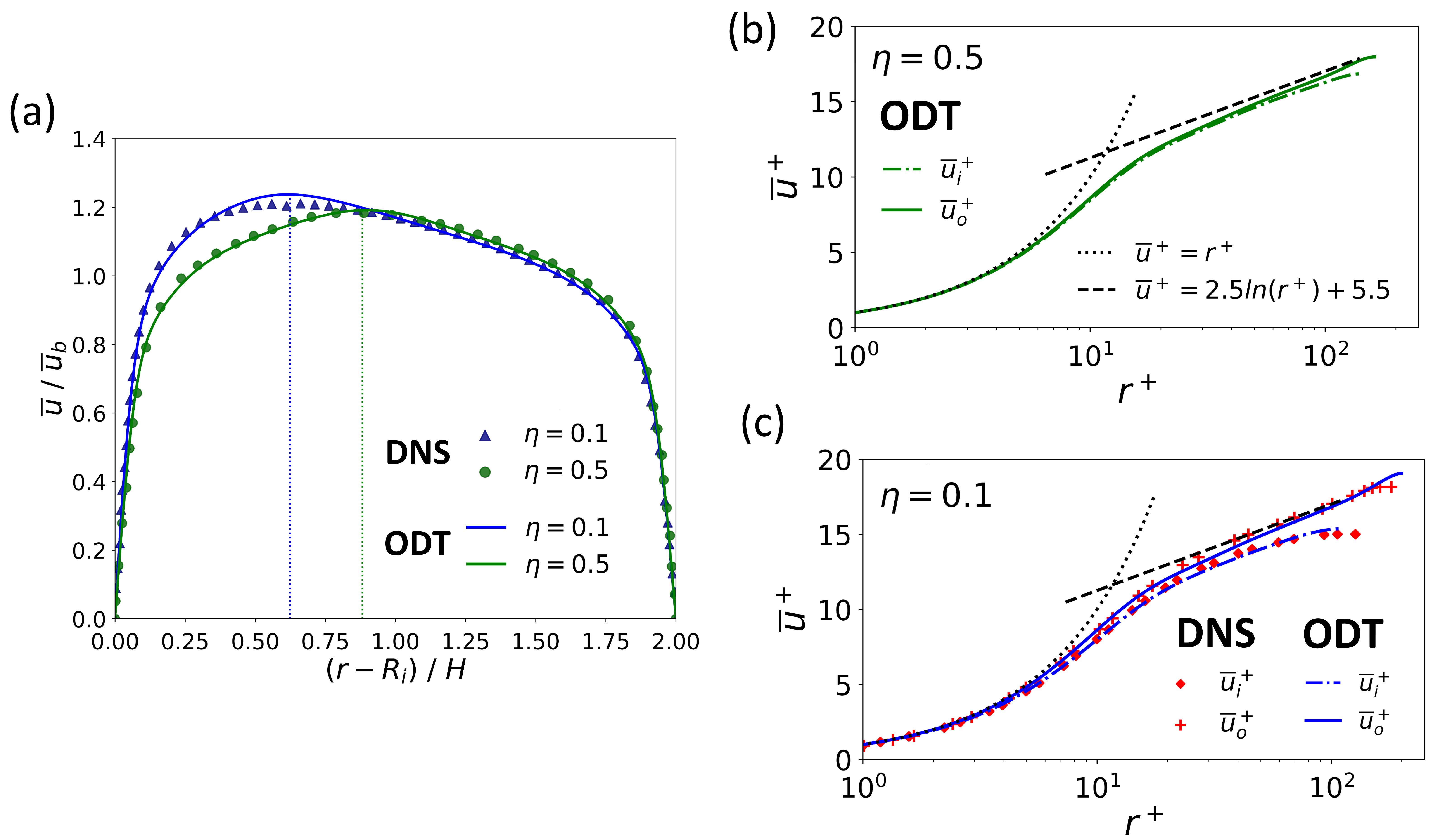}
    \caption{(a) Mean velocity profiles and (b,c) velocity boundary profiles at the inner and outer cylinder walls for radius ratios $\eta=0.5$ and $\eta=0.1$ with a fixed bulk Reynolds number $Re_{D_h}=8900$, compared with reference DNS data~\cite{chung2002}.
    }
    \label{fig:meanVel}
\end{figure}

Figure~\ref{fig:meanVel}(a) presents a comparison of the mean axial velocity profiles for two radius ratios, $\eta=0.1$ and $0.5$, validated against DNS data~\cite{chung2002}. The mean velocity is normalized by the bulk mean velocity, $\overline{u}_\text{b}$. The ODT results show excellent agreement with the DNS data for both radius ratios at a constant bulk Reynolds number of $Re_{D_h}=8900$, demonstrating that the radial asymmetry of the boundary layer is accurately captured. This asymmetry leads to a thicker boundary layer on the outer cylinder surface than on the inner one, with the effect being more significant for $\eta=0.1$ than for $\eta=0.5$. The influence of wall curvature is noticeably stronger near the inner cylindrical wall than at the outer cylindrical wall. This effect becomes even more significant in smaller radius ratios, where the velocity gradient at the inner pipe wall is steeper. This asymmetry is also evident in Table~\ref{table:details}, where the boundary layer thickness on the inner cylinder wall increases with the radius ratio. The normalized boundary layer thicknesses on the inner and outer cylinder walls, denoted as $\delta_{t,\,\text{i}}\,/\,\delta$ and $\delta_{t,\,\text{o}}\,/\,\delta$, respectively, are listed in Table~\ref{table:details}. Additionally, the curvature effect persists even at high Reynolds numbers up to $Re_{D_h}=10^6$. The boundary layer thickness shows only a slight dependence on the Reynolds number, with a marginally smaller $\delta_{t,\,\text{i}}\,/\,\delta$ observed at higher bulk Reynolds numbers. The dotted vertical lines in Figure~\ref{fig:meanVel}(a) indicate the locations of the maximum mean axial velocity, $r_{\text{max}}$, predicted by the ODT model. For $\eta=0.1$ (blue), $r_{\text{max}}=0.615$, and for $\eta=0.5$ (green), $r_{\text{max}}=0.895$. These values correspond to errors of $3.9\%$ and $0.5\%$ relative to the reference DNS data~\cite{chung2002}, respectively. 

%%%%%%%%%%%%%%%%%%%%%%%%%%%%%%%%%%%%%%%%%%%%%%%%%%%%%%%%%%%%%%%%%%%%%%%%%%%%
\subsection{Velocity boundary layers}
Figures~\ref{fig:meanVel}(b) and \ref{fig:meanVel}(c) show the velocity boundary layer profiles at the inner and outer pipe walls for radius ratios $\eta=0.1$ and $0.5$, compared with reference DNS~\cite{chung2002}. It is found that wall curvature does not significantly impact the inner and outer velocity boundary layer profiles in the radius ratio $\eta=0.5$ case. Both normalized profiles align with the master profile, corresponding to the classical law-of-the-wall for pipe flow~\cite{eggels1994}, capturing both the linear viscous sublayer and the logarithmic region. In contrast, significant deviations are evident for the smaller radius ratio of $\eta=0.1$. Here, the inner and outer boundary layers differ substantially, and the inner profile no longer agrees with the law-of-the-wall. This deviation is attributed to the stronger curvature induced by the smaller inner radius, which changes the near-wall classical turbulence structure of the boundary layer. 

\begin{figure}[tbp]
    \centering
    \includegraphics[width=1.0\textwidth]{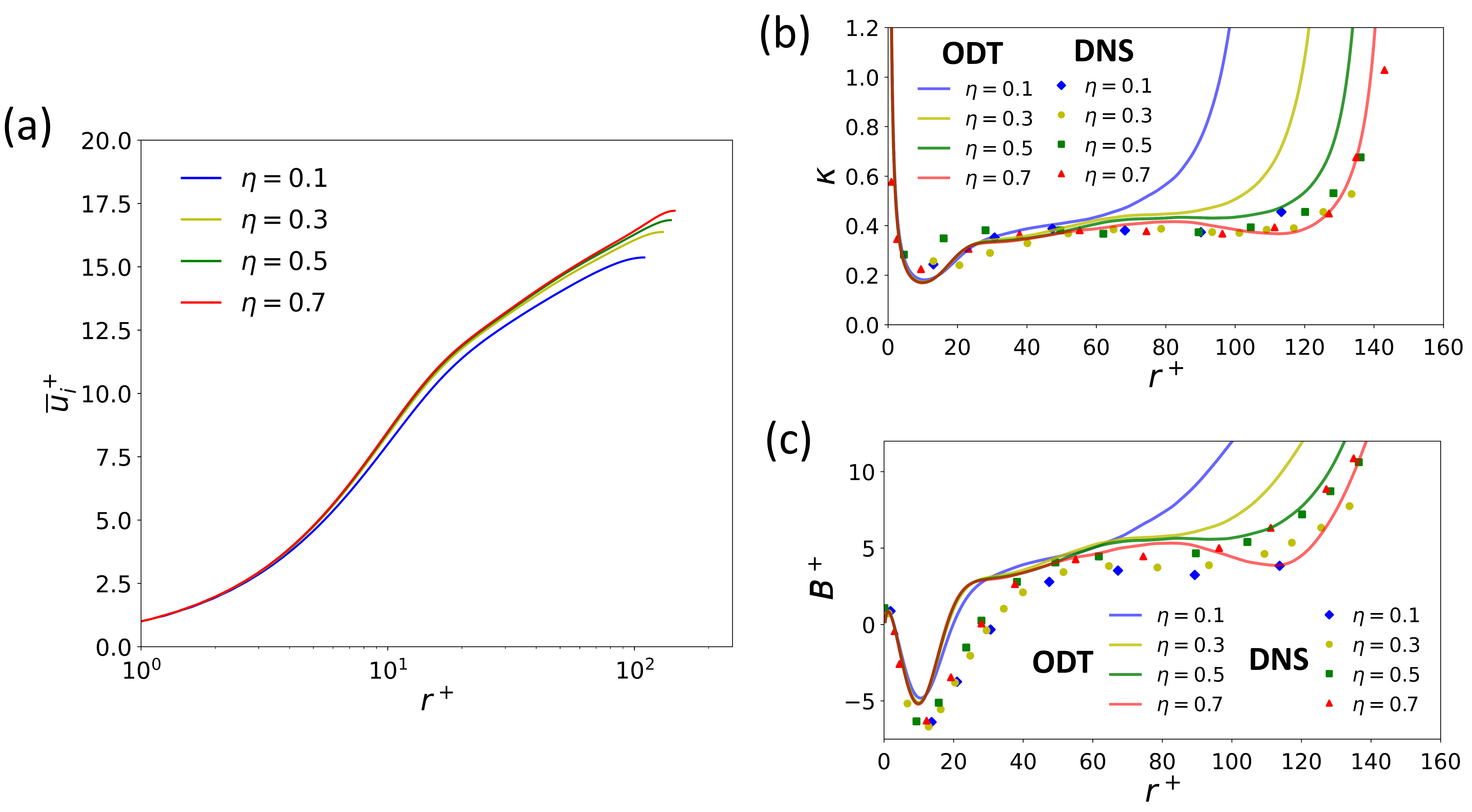}
    \caption{(a) Velocity boundary layer profiles at the inner cylinder wall for radius ratios $\eta = 0.1$, $0.3$, $0.5$, and $0.7$ at a fixed Reynolds number $Re_{D_h} = 8900$. (b,c) Corresponding values of the von Kármán constant $\kappa$ and the intercept $B^+$ for each case, compared with reference DNS data~\cite{bagheri2020}.
    }
    \label{fig:indicatorFunc}
\end{figure}

Figure~\ref{fig:indicatorFunc}(a) further illustrates the effect of wall curvature by comparing the velocity boundary layer at the inner cylinder wall across various radius ratios. The results show a clear trend that as $\eta$ decreases, the extent of the log-law region diminishes, with the case $\eta=0.1$ exhibiting a pronounced departure from the logarithmic behavior. 

To assess the applicability of the classical law-of-the-wall under curved-wall conditions, we apply the logarithmic velocity law in the conventional form,
\begin{equation}
    u^+ = \frac{1}{\kappa} \ln \left( r^+ \right) + B^+ 
     \; \; .
    \label{eq:lawOfWall}
\end{equation}
Here, $\kappa$ is the von K\'arm\'an constant and $B^+$ is an empirical additive constant. These parameters are computed locally from the mean velocity gradient using
\begin{equation}
    \frac{1}{\kappa}=r^+ \frac{\partial u^+}{\partial r^+}
    \; \quad\text{and}\quad
    B^+= u^+ - \frac{1}{\kappa}\ln \left( r^+ \right)
     \; \; .
    \label{eq:vonKarmanCoeff}
\end{equation}

Figures~\ref{fig:indicatorFunc}(b) and \ref{fig:indicatorFunc}(c) compare the extracted values of $\kappa$ and $B^+$ from the inner boundary layer profiles with DNS reference data~\cite{bagheri2020}. While the ODT model does not fully reproduce the exact magnitudes of these constants, it successfully captures the qualitative trend in their variation with wall distance $r^+$. In both cases, a plateau region emerges, where $\kappa$ and $B^+$ are relatively stable values close to the canonical ones for turbulent pipe flow~\cite{eggels1994}. Notably, the extent of this plateau decreases with decreasing $\eta$, highlighting the increasing disruption of the classical boundary layer structure due to curvature effects.

To further quantify the influence of wall curvature on the inner boundary layer, we conduct a velocity boundary layer analysis following the methodology of Boersma \textit{et al.}~\cite{boersma2011}. Based on this framework, the boundary layer near the inner cylindrical wall is decomposed into two distinct regions, the viscous-dominated region and the Reynolds-stress-dominated region. These two regions are discussed in detail in the following sections.

%%%%%%%%%%%%%%%%%%%%%%%%%%%%%%%%%%%%%%%%%%%%%%%%%%%%%%%%%%%%%%%%%%%%%%%%%%%%
\subsubsection{Viscous-dominated region}\label{sec:viscosityRegion}
Assuming the flow is fully developed, all statistical properties of the flow remain unchanged in both the axial and azimuthal directions. As a result, only the steady-state balance equations governing the radial dependence of the statistical moments of the flow variables need to be considered. By applying the Reynolds-averaging procedure to Equation~\eqref{eq:gov_vel}, rearranging the equations, and integrating once over the radial gap (bounded coordinate $r$), we arrive at the shear stress balance equation, which can be expressed as 
\begin{equation}
    \overline{u^\prime v^\prime} - \nu\frac{\partial \overline{u}}{\partial r} = -\frac{r}{2}\frac{1}{\rho}\frac{{\rm d} P}{{\rm d} x}+\frac{C^*}{r}
    \; \quad\text{with}\quad
    C^*=-\frac{1}{\rho}\frac{\tau_\text{o} R_\text{i}-\tau_\text{i} R_\text{o}}{R_\text{i}/R_\text{o} - R_\text{o}/R_\text{i}}
    \; \; .
    \label{eq:gov_balance}
\end{equation}
Here, $C^*$ is an integration constant that is determined by the overall flow solution and the wall shear stresses. The wall shear stress at the inner and outer walls is given by $\tau_\text{i/o} = \mu \left|\left({\mathrm{d}\overline{u}}/{\mathrm{d}r}\right)\right|_{\text{wall,\,i/o}}$, where $\mu$ is the dynamic viscosity. The notation $\tau_\text{i}$ and $\tau_\text{o}$ correspond to the inner and outer walls, respectively. The term $\overline{u^\prime v^\prime}$ represents the statistical contribution of turbulent eddy motions and serves as the model analog of the Reynolds shear stress~\cite{kerstein1999, marten2022}. 

Figures~\ref{fig:BLstress}(b) and \ref{fig:BLstress}(c) show the radial profiles for normalized viscous shear stress, turbulent Reynolds shear stress, and total shear stress for fixed radius ratio $\eta=0.1$, but two different bulk Reynolds numbers $Re_{D_h}=8900$ and $10^6$, respectively. As shown, in the region very close to the wall, where $r^+\leq5$ following~\cite{vonkarman1930}, the flow is dominated by viscous shear stress, meaning $\overline{u^\prime v^\prime}\ll \nu({\rm d}\overline{u}/{\rm d}r)$.  This viscous-dominated region is observed across both low and high Reynolds number cases. For asymptotically small radius ratios, as $\eta\to0$, Equation~\eqref{eq:gov_balance} simplifies to
\begin{equation}
    u_\text{i}^+(r^+)\approx R_\text{i}^+\ln\left( \frac{r^+ + R_\text{i}^+}{R_\text{i}^+} \right)
    \; \quad\text{with}\quad
    R_\text{i}^+=\frac{R_\text{i}\,u_{\tau,\,\text{i}}}{\nu}
    \; \; ,
    \label{eq:visco_dom}
\end{equation}
where $R_\text{i}^+$ is not a constant but a parameter that depends on both the wall geometry and the turbulent flow state, specifically, it is influenced by the curvature radius and the wall-shear stress acting on the cylindrical inner wall. It is noted that, in this formulation, the behavior in the near-wall region deviates from the purely linear behavior described by the classical law of the wall. Instead, as the influence of curvature and turbulence becomes significant, the velocity profile transitions to a logarithmic behavior, reflecting the modified structure of turbulence near curved walls.

%%%%%%%%%%%%%%%%%%%%%%%%%%%%%%%%%%%%%%%%%%%%%%%%%%%%%%%%%%%%%%%%%%%%%%%%%%%%
\subsubsection{Reynolds-stress-dominated region}\label{sec:mixingLengthRegion}
In the region away from the wall, which is referred to as the Reynolds-stress-dominated region, the flow behavior shifts to being dominated by turbulent Reynolds shear stress. In this regime, the viscous contribution becomes negligible compared to the turbulent transport, meaning $\nu({\rm d}\overline{u}/{\rm d}r) \ll \overline{u^\prime v^\prime}$. The influence of the radius is not negligible. For asymptotically small values of the radius ratio, $\eta\to0$, simplification of the integration constant $C_1$ is achieved by assuming $R_\text{o}\gg R_\text{i}$, which preserves radial information. Next, following the methodology of~\cite{boersma2011}, a conventional parameterization of the turbulent eddy viscosity $\nu_t$ is introduced. As suggested in their work, turbulent eddy viscosity can be expressed as
\begin{equation}
    \nu_t=\alpha\sqrt{\frac{\tau_\text{i}}{\rho}}\left( r-R_\text{i}\right)
    \; \; ,
    \label{eq:nut}
\end{equation}
where $\alpha$ is an empirical proportionality factor that has to be estimated with simulation data. Substituting this expression into the momentum balance equation, Equation~\eqref{eq:gov_balance} reduces to
\begin{equation}
    u_\text{i}^+(r^+) \approx \frac{1}{\alpha^+}\ln\left( \frac{r^+}{r^+ + R_\text{i}^+}\right) + D^+
    \; \; ,
    \label{eq:ReyStress_dom}
\end{equation}
where $\alpha^+$ and $D^+$ are parametrization coefficients that need to be determined with flow simulation data. 

\begin{figure}[tbp]
    \centering
    \includegraphics[width=1.0\textwidth]{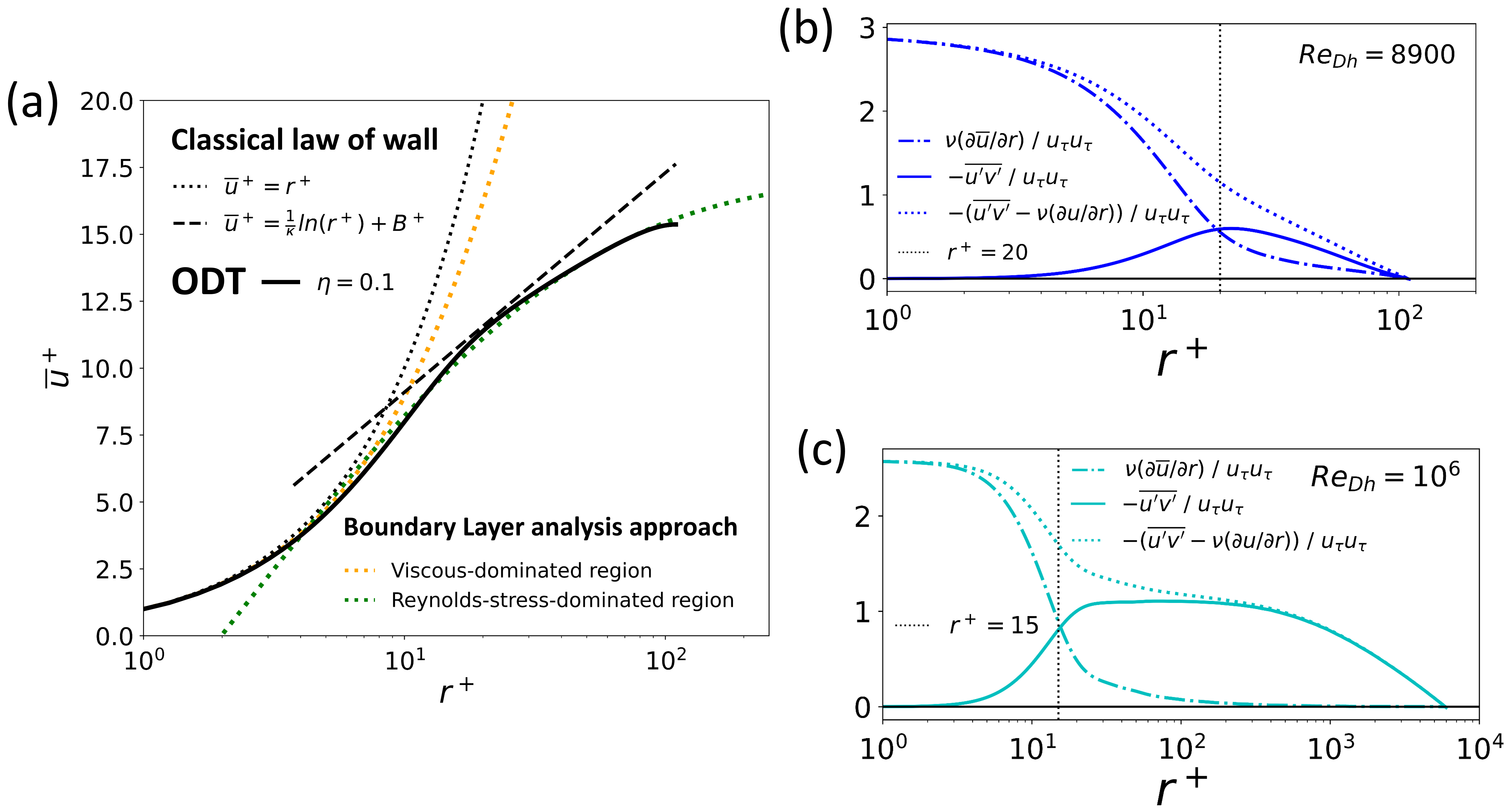}
    \caption{(a) Velocity boundary layer profile at the inner cylinder wall compared with the conventional law of the wall and results from boundary layer analysis. Von K\'arm\'an constant $\kappa=0.28$ and $B^+=0.9$ from reference DNS~\cite{bagheri2020}. Profiles of the distribution of normalized viscous shear stress $\tau_{\text{vis}}/\rho=\nu\left( \partial \overline{u} / \partial r \right)$, turbulent Reynolds shear stress $\tau_{\text{tur}}/\rho=-\overline{u^\prime v^\prime}$, and total shear stress $\tau_{\text{tot}}=\tau_{\text{vis}}+\tau_{\text{tur}}$ at the inner cylinder wall tested for two bulk Reynolds numbers (b) $Re_{D_h}=8900$ and (c) $Re_{D_h}=10^6$ with a fixed radius ratio $\eta=0.1$.
    }
    \label{fig:BLstress}
\end{figure}

Figure~\ref{fig:BLstress}(a) shows the boundary layer at the inner cylinder wall with radius ratio $\eta=0.1$ compared together with conventional and modified wall functions according to Equation~(\ref{eq:visco_dom}) and (\ref{eq:ReyStress_dom}). The conventional wall function is composed of the linear viscous sublayer, $u^+=r^+$, and the conventional logarithmic region, $u^+=\kappa^{-1}\ln(r^+)+B^+$, with parameterizations given by a recent annular pipe flow work from~\cite{bagheri2020}. It is observed that in the viscous-dominated region, Equation~(\ref{eq:visco_dom}) accurately predicts the near-wall velocity profile, with a slightly extended range of validity up to $r^+\le 10$, surpassing the conventional linear wall approximation. In the Reynolds-stress-dominated region, Equation~(\ref{eq:ReyStress_dom}) effectively captures the dominant statistical features of the turbulent boundary layer along the cylindrical inner wall, offering improved accuracy compared to the conventional logarithmic law.

\begin{figure}[tbp]
    \centering
    \includegraphics[width=1.0\textwidth]{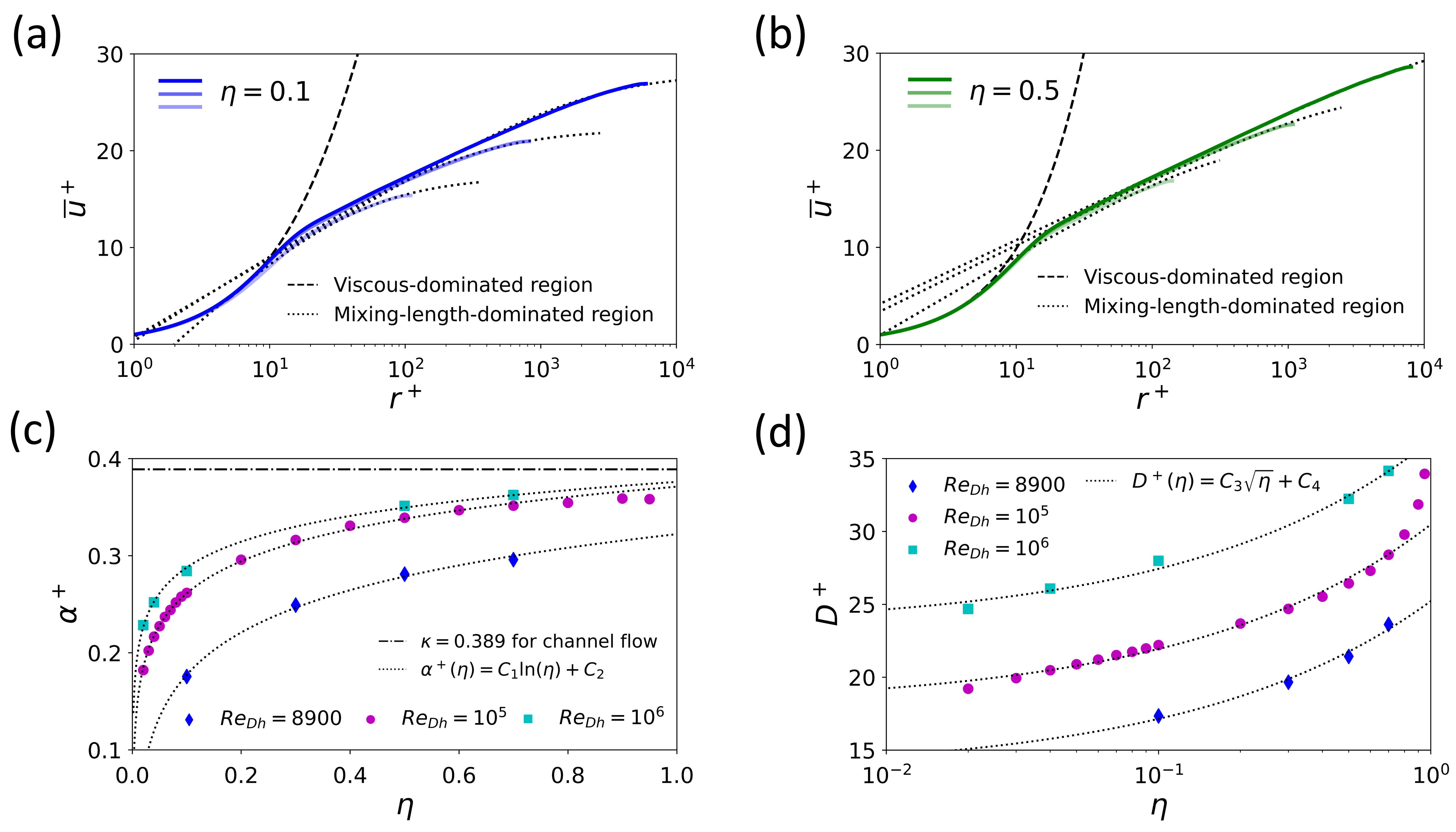}
    \caption{Velocity boundary layer profiles at the inner cylinder wall for radius ratio (a) $\eta=0.1$ and (b) $\eta=0.5$ with various bulk Reynolds numbers $Re_{D_h}=8900, 10^5$ and $10^6$. The profiles of computing coefficients (c) $\alpha^+$ and $D^+$ as a function of radius ratio $\eta$ varying with bulk Reynolds number $Re_{D_h}$.
    }
    \label{fig:BLcoeff}
\end{figure}

Figure~\ref{fig:BLcoeff}(a,b) extend the comparison of velocity boundary layer profiles along the inner cylinder wall to a radius ratio of $\eta=0.5$, across a wide range of bulk Reynolds numbers: $Re_{D_h}=8900,10^5$ and $10^6$. It is found that as the Reynolds number increases, a thicker boundary layer is found, and the curvature effect becomes less significant but still remains noticeable up to $Re_{D_h}=10^6$. Additionally, the modified boundary layer analysis proves applicable even at a radius ratio of $\eta = 0.5$.

Figure~\ref{fig:BLcoeff}(c,d) displays the constant coefficients $\alpha^+$ and $D^+$ from Equation~\ref{eq:ReyStress_dom} across a wide range of radius ratios and Reynolds numbers. Alongside the simulation results, empirical trend lines for both $\alpha^+$ and $D^+$ as functions of the radius ratio $\eta$ are also presented in the figure. These trend lines follow the forms $\alpha^+(\eta)=C_1 \ln(\eta)+C_2$ and $D^+(\eta)=C_3\sqrt{\eta}+C_4$, with the corresponding fitting constants derived from simulation data and summarized in Table~\ref{table:BLcoeff}. The results indicate that as both the radius ratio $\eta$ and Reynolds number increase, the coefficient $\alpha^+$ approaches the von K\'arm\'an constant $\kappa=0.389$ for channel flow~\cite{monty2005}. This trend supports the analogy that in the limiting case of annular pipe flow, where $\eta\rightarrow 1$, the geometry increasingly resembles a planar channel, and the influence of curvature becomes negligible. Furthermore, the diminishing curvature effect with increasing Reynolds number is also clearly observed.
Although the current empirical models capture the general trends of $\alpha^+$ and $D^+$ with respect to $\eta$, further refinement is needed, such as incorporating Reynolds number dependence into the formulations. Nevertheless, the investigation of these coefficients contributes to a better understanding of modified wall functions and may support the development of more accurate wall models in turbulence simulations or enhance the applicability of the present boundary layer analysis framework.

\begin{table}
\centering
\caption{List of the magnitudes of computing coefficients from the equations of boundary layer analysis.}
\label{table:BLcoeff}
\begin{tabular}{ccccc}
\hline 
$Re_{D_h}$ & $C_1$  & $C_2$  & $C_3$   & $C_4$  \\ 
\hline
$8900$    & $0.06$ & $0.32$ & $11.80$ & $13.41$ \\
$10^5$    & $0.05$ & $0.37$ & $12.48$ & $17.99$ \\
$10^6$    & $0.04$ & $0.38$ & $12.86$ & $23.37$ \\ 
\hline   
\end{tabular}
\end{table}

%%%%%%%%%%%%%%%%%%%%%%%%%%%%%%%%%%%%%%%%%%%%%%%%%%%%%%%%%%%%%%%%%%%%%%%%%%%%
\subsection{Reynolds stress components}
\begin{figure}[tbp]
    \centering
    \includegraphics[width=1.0\textwidth]{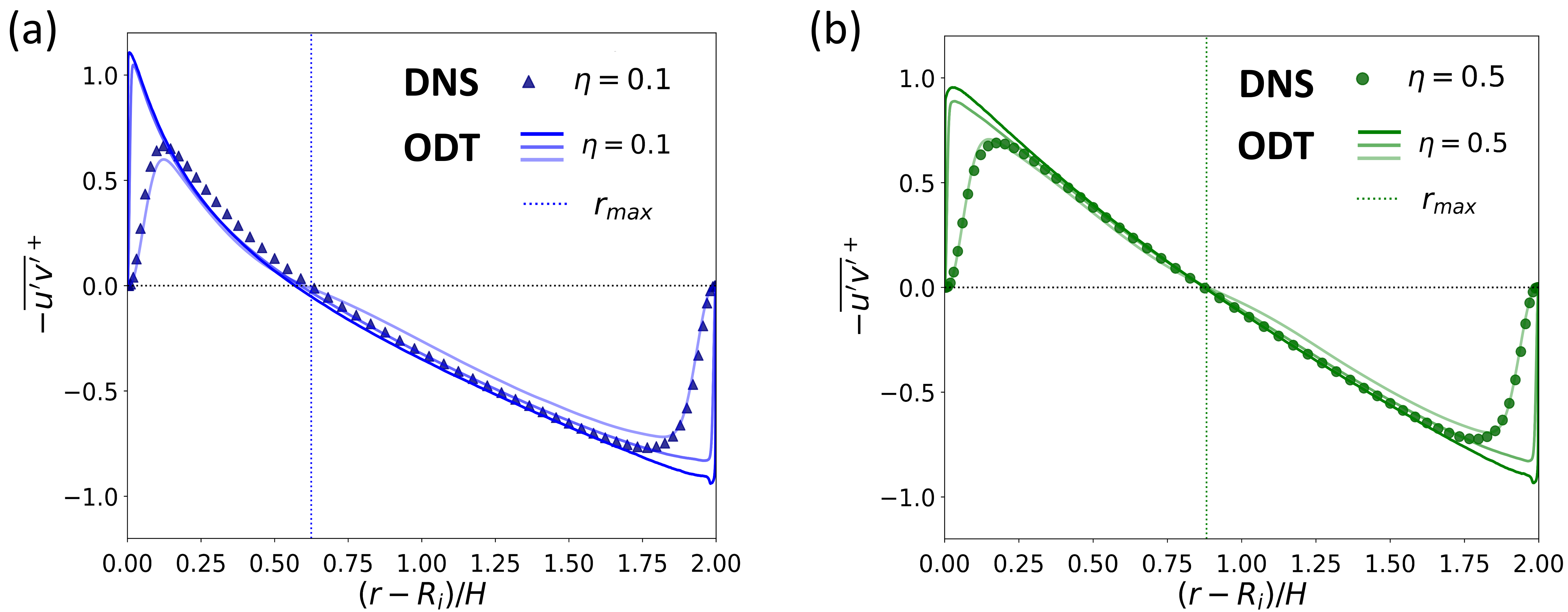}
    \caption{Radial profile of the Reynolds shear stress $-\overline{u^\prime v^\prime}^+$ for radius ratio (a) $\eta=0.1$ and (b) $\eta=0.5$, respectively, and varies with different bulk Reynolds number $Re_{D_h}=8900, 10^5$ and $10^6$ (color from light to dark). Dashed lines denote $r_\text{max}$, the location of the maximum mean axial velocity for $Re_{D_h}=8900$ case. Reference DNS data is from \cite{bagheri2020}.
    }
    \label{fig:RSS}
\end{figure}

Figure~\ref{fig:RSS} presents the Reynolds shear stress (RSS), $-\overline{u^\prime v^\prime}^+$, as a function of wall-normal distance $(r - R_i)/H$ for radius ratios $\eta=0.1$ and $0.5$, from Reynolds number $Re_{D_h}=8900$ to $Re_{D_h}=10^6$. The $Re_{D_h}=8900$ case results show good agreement with the reference DNS data~\cite{bagheri2020}. It is notable that, for the case of $\eta = 0.5$, the zero-crossing point of the Reynolds shear stress (where $-\overline{u^\prime v^\prime}^+=0$) is located at the same position for different Reynolds nymber and coincides with the location of the maximum mean axial velocity, $r_{\text{max}}$. This alignment reflects a more symmetric distribution of turbulence and momentum transport across the annular gap. However, in the case of a smaller radius ratio, $\eta=0.1$, the zero-shear-stress location is found to shift slightly toward the inner pipe wall as the Reynolds number increases. This behavior highlights the increasing asymmetry in the flow field as the curvature becomes more pronounced. The stronger influence of the inner wall curvature alters the balance of turbulent transport and causes the peak in Reynolds shear stress production to shift accordingly.

%%%%% location for zero-crossing point of Reynolds Shear Stress %%%%%
% for eta= 0.1  and Re= 8900  , the zero crossing is at :  0.6039999997777777
% for eta= 0.1  and Re= 100000  , the zero crossing is at :  0.5877999997777777
% for eta= 0.1  and Re= 1000000  , the zero crossing is at :  0.5688599997777778
% for eta= 0.5  and Re= 8900  , the zero crossing is at :  0.8859999999999999
% for eta= 0.5  and Re= 100000  , the zero crossing is at :  0.8792000000000002
% for eta= 0.5  and Re= 1000000  , the zero crossing is at :  0.87408
%%%%%%%%%%%%%%%%%%%%%%%%%%%%%%%%%%%%%%%%%%%%%%%%%%%%%%%%%%%%%%%%%%%%%

\begin{figure}[tbp]
    \centering
    \includegraphics[width=1.0\textwidth]{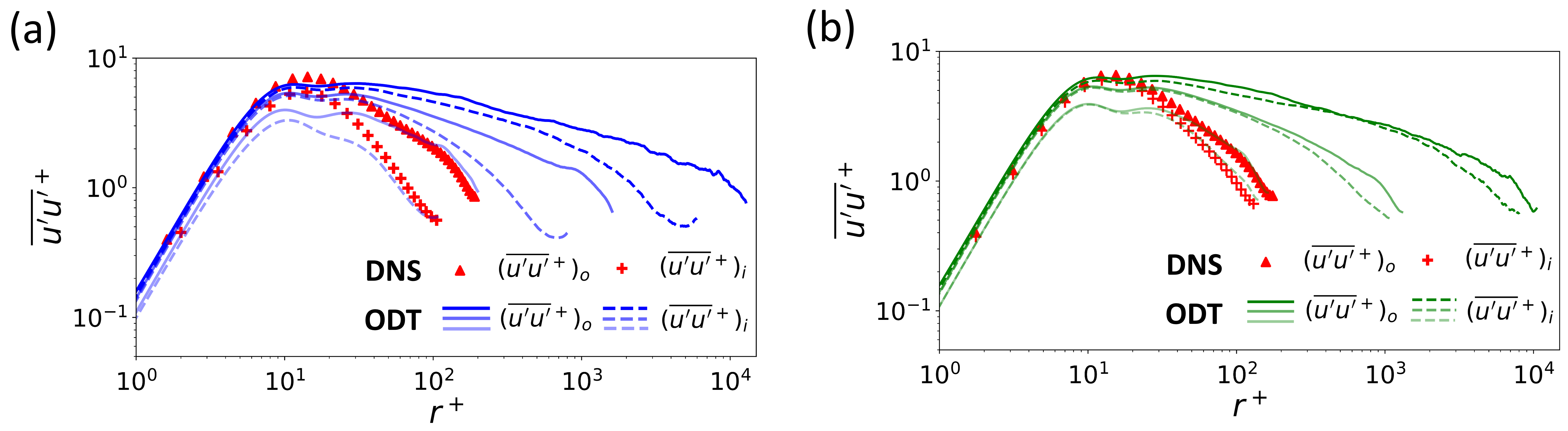}
    \caption{Radial profiles of the axial component of the Reynolds normal stresses in the vicinity of the inner and outer pipe wall, respectively, for bulk Reynolds number $Re_{D_h}=8900, 10^5$ and $10^6$ (color from light to dark) and radius ratio for (a) $\eta=0.1$ and (b) $\eta=0.5$. Reference DNS data is from~\cite{bagheri2020}.
    }
    \label{fig:Urms}
\end{figure}

Figure~\ref{fig:Urms} shows the radial profiles of the axial component of the normal Reynolds stresses, $\overline{u^\prime u^\prime}^+$, near both the inner $(\overline{u^\prime u^\prime}^+)_\text{i}$ and outer $(\overline{u^\prime u^\prime}^+)_\text{o}$ cylinder walls for radius ratios $\eta=0.1$ and $\eta=0.5$, across a range of Reynolds numbers. The results are compared with reference DNS data from~\cite{bagheri2020}. Here, $\overline{u^\prime u^\prime}^+= \left(u_\text{rms}^+\right)^2$ represents the variance (or squared root-mean-square) of the axial velocity fluctuations, serving as a key indicator of turbulence intensity. This comparison aims to evaluate the ability of the model to capture the radial asymmetry of velocity fluctuations across the annular gap. Regions with higher values of $\overline{u^\prime u^\prime}^+$ correspond to zones of more intense turbulence. As expected, the peak in axial velocity variance occurs within the buffer layer, in the range $5<r^+<30$. Furthermore, the results reveal that turbulence intensity is generally higher near the outer wall than the inner wall, particularly in the case of the smaller radius ratio $\eta=0.1$. This difference becomes less pronounced as the radius ratio increases, as observed in the $\eta=0.5$ case. 

In contrast to the reference DNS data~\cite{bagheri2020}, the ODT model significantly underestimates the near-wall peak of turbulent fluctuations within the buffer layer. However, it successfully captures the overall trend of curvature influence. This modeling limitation is known and documented in the literature~\cite{lignell2013,lignell2018,marten2022}. In fact, the ODT model is not specifically optimized for the lower Reynolds number flow. It is primarily intended for highly turbulent regimes, where reference DNS data for annular pipe flows is currently unavailable to support its application in this context. The underestimation of turbulence intensity observed in the ODT results can thus be attributed to finite Reynolds number effects. Previous studies on ODT modeling~\cite{marten2022} have shown that such discrepancies tend to diminish at higher, asymptotic Reynolds numbers, particularly with respect to flux predictions. Therefore, despite the observed limitations at moderate Reynolds numbers, the present validation suggests that the ODT model remains a reliable tool for simulating turbulent annular flows in practical applications, such as heat exchangers, chemical reactors, and gas-cleaning devices, where high Reynolds numbers and strong curvature effects are often encountered.

%%%%%%%%%%%%%%%%%%%%%%%%%%%%%%%%%%%%%%%%%%%%%%%%%%%%%%%%%%%%%%%%%%%%%%%%%%%%
\subsection{Detailed velocity statistics}
%%%%%%%%%%%%%%%%%%%%%%%%%%%%%%%%%%%%%%%%%%%%%%%%%%%%%%%%%%%%%%%%%%%%%%%%%%%%
\subsubsection{Axial velocity fluctuations}
\begin{figure}[tbp]
    \centering
    \includegraphics[width=1.0\textwidth]{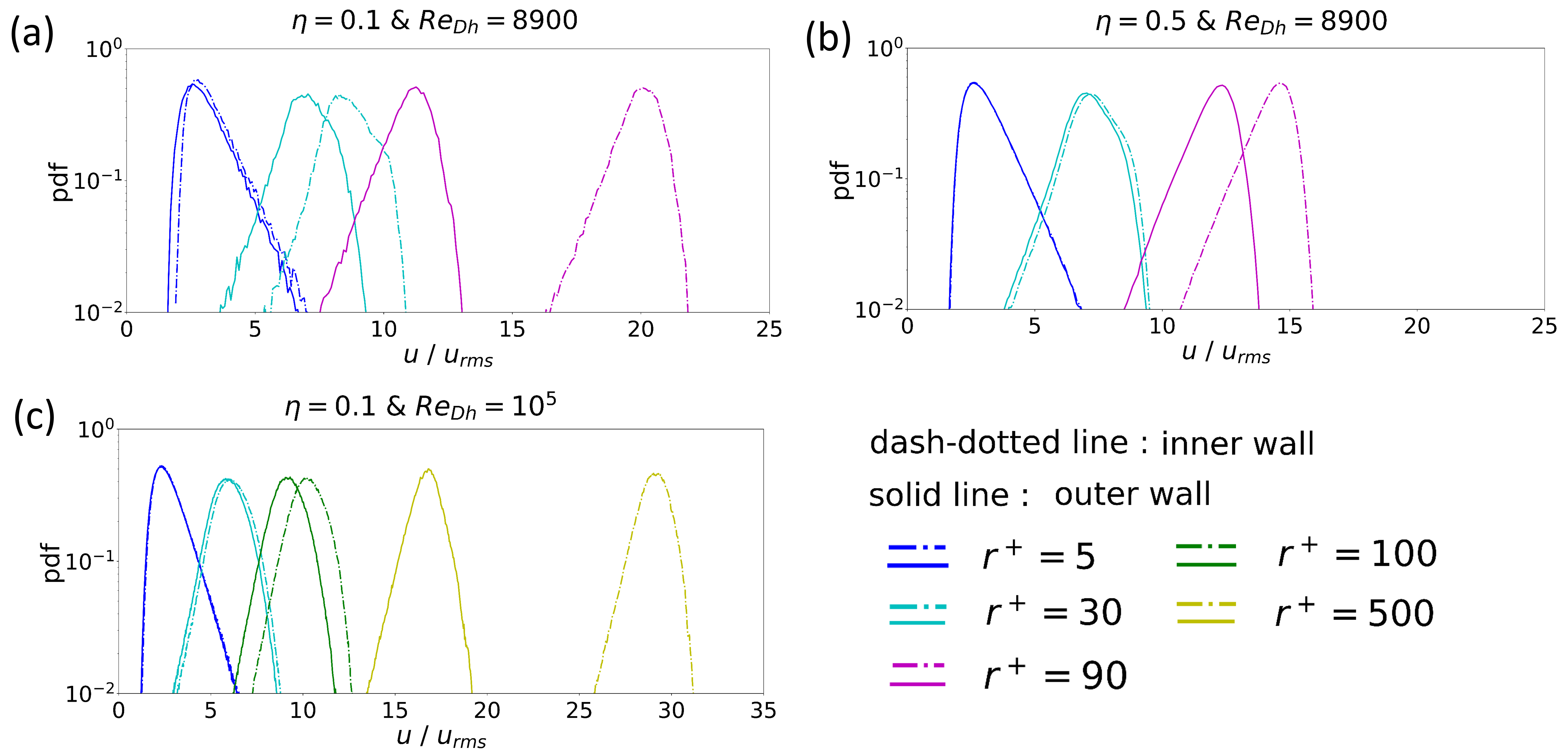}
    \caption{
    Probability density functions (PDFs) of the axial velocity fluctuations $u^\prime$ normalized by the RMS $u^\prime_\text{rms}$ at $r^+=5$ in the viscous sublayer and $r^+=30, 90, 100, 500$ in the turbulent log layer on the inner and outer pipe walls for radius ratio (a) $\eta=0.1$ and (b) $\eta=0.5$ and bulk Reynolds number $Re_{D_h}=8900$ and (c) $Re_{D_h}=10^5$. 
    }
    \label{fig:PDF}
\end{figure}

To better understand the variation in axial velocity fluctuation statistics with distance from the wall, we analyze the probability density function (PDF) of the fluctuating axial velocity over the simulation time. Specific wall-normal positions are selected on both the inner and outer cylinder walls: $r^+=5$, representing the viscous sublayer, and $r^+\geq30$, corresponding to the Reynolds-stress-dominated region. Figure~\ref{fig:PDF} compares the PDFs of the normalized fluctuating axial velocity, $u/u_\text{rms}$, at these selected $r^+$ values for both sides of the annular gap.

The results show that the PDFs generally follow a Gaussian-like distribution. However, as the wall-normal distance increases, the peaks of the PDFs shift toward higher values of $u/u_\text{rms}$, leading to the shape of the distribution profiles. While the overall shape of the PDFs remains similar between the inner and outer walls, differences in the magnitude of normalized fluctuations become more apparent further away from the wall. In particular, the discrepancy between the inner and outer wall profiles grows with increasing $r^+$, highlighting a stronger curvature effect on axial velocity fluctuations in the outer region of the flow.This increase is more pronounced in the $\eta=0.1$ case (Figure~\ref{fig:PDF}(a)) than in the $\eta=0.5$ case (Figure~\ref{fig:PDF}(b)). Additionally, it is observed that the PDF on the inner wall tends to exhibit higher values of $u/u_\text{rms}$ compared to the outer wall. This suggests that the asymmetric geometry of the annular pipe leads to uneven turbulent energy distribution between the two walls, particularly under moderate Reynolds number conditions. A comparison between the low Reynolds number case (Figure~\ref{fig:PDF}(a)) and the high Reynolds number case (Figure~\ref{fig:PDF}(c)) further illustrates the effect of velocity fluctuations on curvature influence. At higher Reynolds numbers, the differences in $u/u_\text{rms}$ between the inner and outer walls at the same $r^+$ location become smaller. This indicates that in strongly turbulent regimes, the influence of curvature on fluctuation statistics diminishes, but it is still visible in the outer region of the flow.

%%%%%%%%%%%%%%%%%%%%%%%%%%%%%%%%%%%%%%%%%%%%%%%%%%%%%%%%%%%%%%%%%%%%%%%%%%%%
\subsubsection{Turbulent eddy event statistics}
The ODT model simulates eddy turbulence by stochastic sampling of eddy events.
In boundary-layer-type flows, near-wall shear is the main driver for turbulence production due to shear-available energy. 
The ensemble of radial locations and sizes of the sampled eddy events therefore encodes surrogate information about the turbulence properties, that is, the turbulence activity and the participating scales at any selected wall distance.
The ensemble of sampled eddy events represents discrete labeled data that is here used to gain insight into the composition of the turbulent flow in the vicinity of the inner and outer cylinder wall, respectively. 
Based on the eddy midpoint location $r_\text{m}$ and the eddy size $\ell$, the eddies are classified based on a simple geometric rule as \textit{attached eddies} and \textit{detached eddies}. Attached eddies are defined by having the edge closest to the wall within the distance of the eddy size so that the midpoint lies within the range $\ell/2 < |r_\text{m}-R_\text{i/o}| < 3\ell/2$. Detached eddies are all other eddies, and these have their midpoint at wall distance $|r_\text{m}-R_\text{i/o}| > 3\ell/2$. These are the only geometrically possible configurations so that the sorting criterion, expressed in friction units, simplifies to $r_\text{m}^+\gtrless3\ell^+/2$ for detached and attached eddies that are in general closer to the inner or outer wall, respectively. 

\begin{figure}[tbp]
    \centering
    \includegraphics[width=1.0\textwidth]{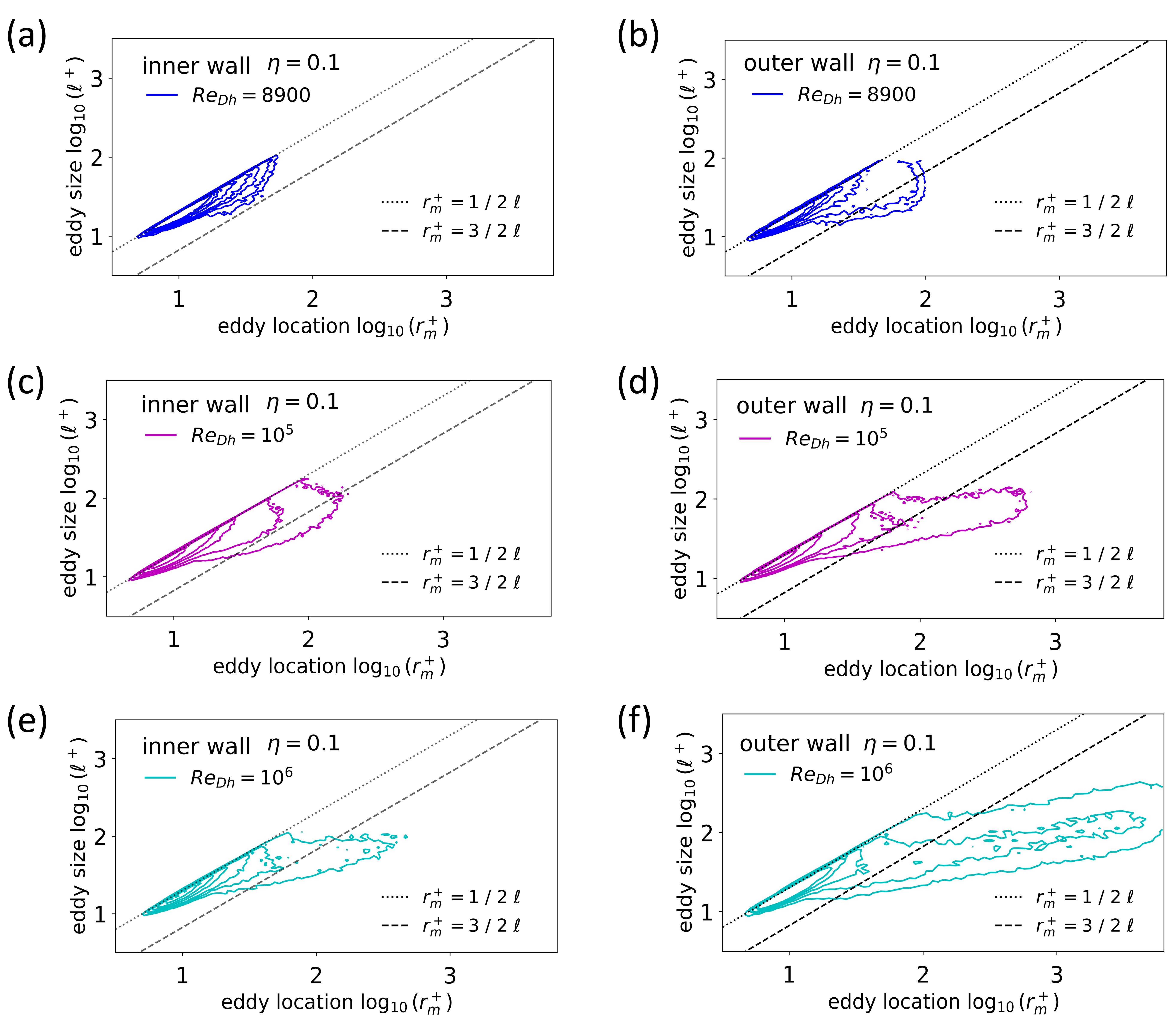}
    \caption{Joint probability density functions (jPDFs) of the eddy event's normalized midpoint location $r_\text{m}^+$ and eddy size $\ell^+$ at the inner and outer cylinder wall, respectively, for various  Reynolds numbers $Re_{D_h}$, but fixed radius ratio $\eta=0.1$. 
    }
    \label{fig:Eddy}
\end{figure}

Figure~\ref{fig:Eddy} presents the joint probability density functions (jPDFs) of eddy events' radial locations and sizes, comparing three different Reynolds numbers at a fixed radius ratio of $\eta=0.1$ for which curvature effects are strong. 
It is found that, near the inner cylinder wall, eddies tend to be smaller in size when located closer to the wall, with sizes increasing farther away. It is also observed that attached eddies occupy the inner cylinder region, whereas the majority of detached eddies are found in the outer cylinder region. As the Reynolds number increases, the number of detached eddies in the outer layer becomes more significant, indicating enhanced turbulent activity in that region. This trend indicates that detached eddies primarily form downstream of the location where the maximum axial velocity occurs. The increase in detached eddy activity with Reynolds number suggests enhanced turbulent mixing and more pronounced flow separation effects at higher Reynolds numbers, particularly in regions farther from the inner wall.

%%%%%%%%%%%%%%%%%%%%%%%%%%%%%%%%%%%%%%%%%%%%%%%%%%%%%%%%%%%%%%%%%%%%%%%%%%%%%
\section{Conclusions}\label{sec:conclusions}
In this study, a stochastic turbulence modeling approach is applied to simulate pressure-driven turbulent flow in a concentric annular pipe, with a focus on investigating the effects of spanwise curvature and Reynolds number. The numerical simulations were conducted using the one-dimensional turbulence (ODT) model, which represents turbulent transport through stochastic eddy events. The model is validated against available reference direct numerical simulation (DNS) data~\cite{chung2002, bagheri2020} and is employed as a cost-efficient tool to resolve turbulent boundary layer dynamics at high Reynolds numbers. Despite the radically reduced dimensionality, the ODT model incorporates the boundary layer approximation and retains full resolution of radial transport processes.

The model calibration and validation was performed by comparing the ODT results with DNS data at a bulk Reynolds number of $Re_{D_h}=8900$, focusing on key flow quantities, including mean velocity profiles, velocity boundary layers on both the cylindrical inner and outer wall, Reynolds shear stress, and axial Reynolds stress components. The ODT model demonstrates good agreement in reproducing first-order flow statistics. However, it tends to underestimate axial velocity fluctuations in the buffer layer, which is a well-known modeling artifact~\cite{lignell2013} that persists for a radial domain and flows at low radius ratios and high Reynolds numbers. Importantly, this underestimation does not necessarily imply inaccuracies in radial fluxes. This is because turbulent transport and fluctuation variance are the independent results of the spatial triplet-mapping operations used in the map-based advection modeling of turbulent transport. Hence, the ability of the model to capture radial asymmetry and Reynolds shear stress distributions indicates that it maintains a physically meaningful statistical representation of turbulence-driven transport that enables predictions within a dimensionally reduced framework.

The model results in turn reveal that spanwise curvature introduces pronounced asymmetry in the mean velocity profile and affects the relative boundary layer thicknesses on the inner and outer walls. To improve the prediction of velocity profiles near the inner wall, a theoretical framework has been developed, distinguishing between the viscous-dominated region and the Reynolds-stress-dominated region. By applying classical boundary layer and mixing-length theories, this analysis enables accurate reconstruction of velocity profiles across the boundary layer, laying the groundwork for curvature-modified wall functions and parameterized model coefficients, which will be investigated in future work.

\section*{Acknowledgement}
This research is supported by the Federal Ministry of Research, Technology and Space and the State of Brandenburg within the framework of the joint project EIZ: Energy Innovation Center (project numbers 85056897 and 03SF0693A).

\bibliographystyle{elsarticle-num}

\bibliography{literature}

\end{document}